\documentclass[12pt]{article}
\usepackage[utf8]{inputenc}
\usepackage[T1]{fontenc}
\usepackage{amsmath,amsthm,amsfonts,amssymb,url,enumitem,bm}
\usepackage{graphicx}
\usepackage{xcolor,soul}
\usepackage[noabbrev,capitalize]{cleveref}\crefname{equation}{}{} 
\usepackage{breakurl}
\usepackage{eucal}
\usepackage{scalerel}
\usepackage{algorithm}
\usepackage{algpseudocode}

\newcommand{\xx}{\bm{x}}
\newcommand{\cc}{\bm{c}}
\renewcommand{\ll}{\bm{\lambda}}

\newcommand{\later}[1]{}
\newcommand{\wt}{\mathit{wt}}
\DeclareMathAlphabet{\mathpzc}{OT1}{pzc}{m}{it}
\usepackage{authblk}

\title{Quantum annealing with inequality constraints: the set cover problem}

\author[$\dagger$,$\star$]{Hristo N.\ Djidjev}
\affil[$\dagger$]{
	Institute of Information and Communication Technologies, Bulgarian Academy of Sciences, Sofia, Bulgaria}
\affil[$\star$]{
	Los Alamos National Laboratory, Los Alamos, NM 87545, USA
}

\begin{document}
	
	\later{
		\author[1,2,*]{Hristo N.\ Djidjev}
		\affil[1]{Institute of Information and Communication Technologies, Bulgarian Academy of Sciences, Sofia, Bulgaria}
		\affil[2]{Los Alamos National Laboratory CCS-3 Information Sciences, Los Alamos, NM 87545, USA}
		\affil[*]{djidjev@parallel.bas.bg}
	}
	
	\maketitle

	\begin{abstract}
		This paper presents two novel approaches for solving the set cover problem (SCP) with multiple inequality constraints on quantum annealers. The first method uses the augmented Lagrangian approach to represent the constraints, while the second method employs a higher-order binary optimization (HUBO) formulation. Our experimental analysis demonstrate that both approaches outperform the standard approach with slack variables for solving problems with inequality constraints on  D-Wave quantum annealers. The results show that the augmented Lagrangian method can be successfully used to implement a large number of inequality constraints, making it applicable to a wide range of constrained problems beyond the SCP. The HUBO formulation performs slightly better than the augmented Lagrangian method in solving the SCP, but it is less scalable in terms of embeddability in the quantum chip. These findings could impact the use of quantum annealers for solving constrained optimization problems. 
	\end{abstract}
	
	\vspace{2pc}

\section{Introduction}
\subsection{Quantum annealing}
Quantum annealers, such as those produced by D-Wave Systems Inc., leverage quantum mechanical phenomena, including entanglement and tunneling, to tackle NP-hard optimization problems that are inherently challenging for classical computers. While D-Wave's machines are currently the most powerful quantum devices available, boasting over 5000 qubits, the state of Noisy Intermediate-Scale Quantum (NISQ) technology is still limited in its abilities and cannot yet outperform classical computers in solving practical problems due to high levels of noise and decoherence. However, each new generation of D-Wave quantum annealers is improving hardware to reduce noise and increase coherence time, enabling the machines to accurately solve progressively more complex problems.

To solve a problem on a quantum annealer, it has to be formulated as a problem of the type
\begin{equation}
	\mbox{minimize} \; \mathit{Is}(\xx) = \sum_{i<j}J_{ij}x_ix_j+\sum_ih_ix_i,\label{eq:Ising1}
\end{equation}
where $\xx=\{x_1,\dots,x_n\}$, $J_{ij}$ and $h_i$ are real numbers, and variables $x_i$ are either in $\{-1,1\}$, in which case the problem is called an \textit{Ising problem},
or in $\{0,1\}$, when the formulation is called a \textit{quadratic unconstrained binary optimization} (\textit{QUBO}) problem. The two representations are equivalent and can be easily converted into each other by using a linear variable transformation.

Problem \cref{eq:Ising1}, which is a quadratic function of the variables $x_i$, is known to be NP-hard \cite{Barahona1982} and many important optimization problems can be easily formulated as Ising or QUBO problems \cite{Lucas2014}. Such formulations are usually constructed in two steps. In the first step, the problem of interest is stated as a 0-1 quadratic programming problem, i.e., a problem to minimize a quadratic form of $n$ binary variables subject to linear equality or inequality constraints. 
In the second step, that constrained problem is converted into an unconstrained one, which is necessary since problem \cref{eq:Ising1} formulation doesn't allow constraints. Next we discuss methods to convert constrained problems into unconstrained ones.

\subsection{Handling constraints}
The \textit{penalty method} is the most commonly employed technique for dealing with constrained problems, in which the constraints are included in the objective function as penalty terms. For instance, a constraint ${\bm{a}}^\intercal\xx=b$ can be added to the objective as a penalty term $P(\xx)=\mu(\bm{a}^\intercal\xx-b)^2$, where $\mu>0$ is a \textit{penalty constant}. If $\xx$ satisfies the constraint, then $P(\xx)=0$ and the penalty term doesn't change the value of the objective. But if $P(\xx)\neq 0$, then $P(\xx)> 0$ and if the constant $\mu$ is chosen large enough, then the penalty term will prevent $\xx$ to be a minimum of the combined objective. 

If we have an inequality constraint, we can first convert it into an equality one and then proceed as described above. For instance, a constraint ${\bm{a}}^\intercal\xx\leq b$ can be represented as ${\bm{a}}^\intercal\xx+d= b$, where $d\geq 0$ is a new slack variable. But since problem \cref{eq:Ising1} accept only binary variables and $d$ can potentially be as large as $b$, integer variable $d$ has to be encoded using $\lfloor \log b \rfloor +1$ binary variables. This may lead to a large number of new variables, especially if the problem has multiple inequality constraints. But with the size of the problem increasing, the limited availability of qubits and the reduced precision of current generation quantum devices can pose significant challenges \cite{Pelofske2019mc}. 

Another issue with the penalty method is that a large penalty constant $\mu$ can significantly impact the accuracy of quantum annealing. This is because all $J_{ij}$ and $h_i$ coefficients from \cref{eq:Ising1} are normalized before being submitted to the annealer to satisfy hardware-imposed restrictions of the D-Wave device. Consequently, a large value for $\mu$ leads to some coefficients of the resulting problem being very small in absolute value. This poses a challenge due to the analog nature of the quantum device and the finite precision of the digital-to-analog converter, resulting in further degradation of the accuracy of the quantum annealing.  

\subsection{The set cover problem}
The problem we are considering in this paper, whose formulation involves dealing with multiple inequality constraints, is the \textit{set cover problem (SCP)}. SCP  is a classical optimization problem with multiple applications including in scheduling, resource allocation, logistics, and bioinformatics \cite{vemuganti1998applications}. The weighted version of the problem is, given a set $U$ of $n$ elements $\{1,\dots,n\}$ and a set $M$ of $m\geq 2$ sets $S_i\subset U$ with positive weights $\wt_i$ such that $\bigcup_{S\in M} S=U$, to find $M^*\subseteq M$ for which $$\bigcup_{S_i\in M^*} S_i=U$$ that minimizes $$\sum_{S_i\in M^*}\wt_i.$$ 

In its unweighted version, all weights are one, and both versions are NP-hard \cite{Karp72_NP}. In this paper we consider the weighted version of the SCP. 

\subsection{Our objectives}
In this work, we propose two new approaches for more accurate constraint handling in solving larger SCP problems on current quantum annealers. The first approach, described in the next section, uses the augmented Lagrangian method to represent inequality constraints, as an alternative to the penalty method. The second approach formulates the SCP as a HUBO problem, which is similar to the QUBO formulation presented in \cref{eq:Ising1} but permits higher degree monomials. These methods are intended to address the limitations imposed by the restricted number of qubits and the reduced precision of quantum annealers when handling a large number of variables. 

This work makes three main contributions. Firstly, we demonstrate that the augmented Lagrangian method can effectively handle a large number of inequality constraints when solving constrained problems with quantum annealing. While our application focuses on the SCP, this approach has broad applicability to other problem types. Secondly, we show that a HUBO representation can be leveraged to solve the SCP on a quantum annealer. This method can be extended to other problems with inequality constraints, although the problem must have certain structural characteristics. Lastly, our proposed methods produce QUBO or HUBO problems with only $m$ variables, and our experimental results indicate that the current D-Wave Advantage machines can solve problems with up to 400 sets (variables).

The paper is structured as follows. In \cref{sec:previous}, we give a brief literature review of related results. In \cref{sec:methods}, we provide a brief introduction to the augmented Lagrangian method and the HUBO formulation, and detail our proposed algorithms for handling constraints in quantum annealing. In \cref{sec:results}, we present the experimental results, including a comparison of the accuracies achieved by our methods and those achieved by the penalty method. Finally, we conclude the paper with a summary of our findings and suggestions for future research directions.

\section{Previous work}\label{sec:previous}
Given the set cover problem's significance, many classical heuristic algorithms have been proposed for its solution. 
For the unweighted version of the problem, Johnson \cite{Johnson74} and Lov\'{a}sz \cite{lovasz1975ratio} proposed algorithms that can find set covers with cost at most $1 + \ln d$ times the optimal one, where $d$ is the maximum cardinality of sets in $U$, while Chv\'{a}tal \cite{chvatal1979greedy} generalized their result for the weighted version. 
Feige \cite{feige1998threshold} showed that the set cover cannot be approximated within a factor of $\ln n$.
A review of several algorithms for the SCP and comparison of their practical performances is given in \cite{caprara2000algorithms}. 

There is much less work reported in the literature on quantum algorithms for the SCP. 
Lucas \cite{Lucas2014} describes a QUBO formulation for the SCP, among other NP-hard problems, but does not implement it on a quantum annealer. The number of variables for his formulation is $m+n(\log m+1)$.
For the related problem of \textit{minimum vertex cover}, Pelofske et al. \cite{pelofske2019solving} design a quantum annealing algorithm that can deal with problems that are too large to fit onto the quantum hardware by decomposing them into smaller subproblems. Zhang et al. \cite{zhang2022applying} propose quantum approximate optimization algorithm (QAOA), which uses a model for quantum computing different from quantum annealing, for the minimum vertex cover problem and apply it to graphs of ten vertices.
However, the minimum vertex cover problem is simpler than set cover in the sense that its standard formulation involves a single equality constraint and no inequality ones. Cao et al. \cite{cao2016solving} consider the \textit{set cover with pairs} problem, introduced in \cite{hassin2005set}, and propose a QUBO formulation for it that uses $O(nm^2)$ binary variables. They run a quantum annealing simulator on instances requiring no more than 19 variables, and also solve the QUBO problems using simulated annealing on the same set of instances.

The augmented Lagrangian method (ALM) was introduced in 1969 by Hestenes and Powell \cite{hestenes1969multiplier,powell1969method}  and has been widely studied in the field of optimization. Different variants of the method have been applied to a wide range of problems, including quadratic programming \cite{schittkowski1983convergence}, nonlinear programming \cite{di1979new}, and convex optimization \cite{iusem1999augmented}. In quantum computing,  ALM has been used in \cite{schade2022parallel} for solving the quantum-chemical ground-state energy problem on a gate-based quantum computers. Yonaga et al. \cite{yonaga2020solving} use the alternating direction method of multipliers, a variant of ALM, to solve the quadratic knapsack problem. Djidjev \cite{djidjev2023logical} applies ALM for representing logical qubits in quantum annealers.

We also use higher order binary optimization (HUBO) problem formulations, which have been the subject of intense research. Several works have focused on quadratization, or reduction of the HUBO to a quadratic form. 
Kolmogorov and Zabih \cite{kolmogorov2004energy} and Freedman and Drineas \cite{freedman2005energy} show how to quadratize any monomial with a negative coefficient, regardless of its degree, by introducing a single auxiliary variable. Ishikawa \cite {ishikawa2010transformation} propose a method that results in a more efficient quadratization for positive monomials, utilizing approximately half the number of variables compared to previous techniques. 
Methods that don't introduce auxiliary variables but instead enumerate assignments to a small subset of the variables in order to reduce the degree include using deductions \cite{tanburn2015reducing} and excludable local configurations \cite{ishikawa2014higher}.
Boros and Gruber \cite{boros2014quadratization} review the previous work on quadratization and propose new techniques. HUBO formulations have been used for quantum annealing by Pelofske et al. for Boolean tensor networks \cite{pelofske2022BTN}, by Mato for molecule unfolding, and Jun for prime factorization \cite{jun2023hubo}.

\section{Methods}\label{sec:methods}
\subsection{Using slack variables}\label{sec:slack}
The standard approach to deal with the inequalities of the SCP is using slack variables in order to convert them to equalities and then use the quadratic penalty method to incorporate the resulting equalities into the objective function  \cite{Lucas2014}. Specifically, we define a binary variable $x_i\in\{0,1\}$ for each set $S_i\in M$ that indicates whether the set is included in the final solution or not. Then the objective function to minimize is
$$Q_A=\sum_{i=1}^m x_i \wt_i.$$
If we define $\sigma_{i}=\{j~|~S_j\ni\, i\}$, the constraint that at least one of the selected sets covers element $i$ is $\sum_{j\in\sigma_i}x_j\geq 1$. We convert this into the equality $$\sum_{j\in\sigma_i}x_j = d_i+1,$$ where $d_i$ is a new integer variable in $[0,m-1]$. We encode $d_i$ using $k=\lfloor\log (m-1)+1\rfloor$ binary variables $x_{i,\alpha}$. Then the QUBO encoding all the constraints is
\begin{equation}
	Q_B=\sum_{i=1}^m\bigg(\big(\sum_{j\in\sigma_i}x_j- \sum_{\alpha=0}^k 2^\alpha x_{i,\alpha} - 1\big)^2\bigg).\label{eq:SV}
\end{equation}
Finally, we combine $Q_A$ and $Q_B$ into a single QUBO
$$Q=Q_A+\mu\,Q_B,$$
where $\mu$ is a constant satisfying $\mu>\max\{\wt_i\}$ \cite{Lucas2014}. Although the penalty constant $\mu$ itself is usually not large, the issue is with the coefficients $2^\alpha$, which may become as large as $m$.  In the next two sections, we describe the proposed new approaches.

\subsection{Augmented Lagrangian version}
\subsubsection{The general method}	
The augmented Lagrangian method (ALM) for solving constrained problems combines the penalty method, used in the previous subsection, with the method of the Lagrangian multipliers. Specifically, in the case of inequalities, all inequality constraints of type $c_i(\xx)={\bm{a}_i}^\intercal\xx- b_i\leq 0$, $i=1,\dots,n$, can be included into the objective as an additive term 
$$\ll^\intercal\cc(\xx)+\dfrac{\mu}{2}||\cc(\xx)||^2 = \sum_{i=1}^n \big(\lambda_ic_i(\xx) + \dfrac{\mu}{2}||c_i(\xx)||^2\big), $$
where the coefficients $\lambda_i$ are called \textit{Lagrangian multipliers}, $\ll=\{\lambda_1,\dots,\lambda_n\}$, and $\cc=\{c_1,\dots,c_n\}$. Coefficients $\ll$ and $\mu$ are estimated using an iterative procedure as described in \cref{alg:AL}. Note that a version with both equality and inequality constraints is possible.
\begin{algorithm}
	\caption{Augmented Lagrangian method for inequalities}\label{alg:AL}
	\begin{algorithmic}[1] 
		\Require initial $\mu > 0$, initial $\ll$, increase factor $\rho>1$
		\Ensure final $\mu$,  $\ll$ for use in the augmented Lagrangian function 
		\Repeat
		\State $\xx \gets \arg\min_{\scaleto{\xx}{4.6pt}\in \{0,1\}^n}\,\big(f(\xx)+\ll^T\, \cc(\xx) + \dfrac{\mu}{2}\,||\cc(\xx)||^2\,\big)$\par
		\hskip\algorithmicindent\Comment{Can solve on a quantum annealer}
		\For {$i=1,\dots,n$}
			\If {$c_i(\xx)> 0$}
				\State $\lambda_i \gets \lambda_i + \mu c_i(\xx)$
			\EndIf
		\EndFor
		\State $\mu \gets \rho\mu$ 
		\Until{$\cc(\xx)\leq \bm{0}$ or iteration limit reached} \par
		\hskip\algorithmicindent\Comment{Other stopping criteria possible}
	\end{algorithmic}
\end{algorithm}

In the next subsection we apply the method to the SCP.

\subsubsection{Applying the AL method to the SCP}

First, we formulate the SCP problem as a 0$-$1 linear program with constraints and then we apply the AL method to get rid of the inequalities. As in  \cref{sec:slack}, we define a binary variable $x_i$ for each $i\in[1,m]$ such that $x_i=1$, if subset $S_i$ is selected for the cover, or $x_i=0$, otherwise. Then the SCP can be formulated as
\begin{gather}
	\underset{x_i} { \mbox{minimize}}\quad \sum_{i=1}^mx_i\wt_i \label{eq:SCP1}\\
	\mbox{subject to}\quad \sum_{j\in\sigma_i}x_j\geq 1, \; x_j\in\{0,1\},\;  i=1,...,n. \label{eq:SCP2}
\end{gather}

The corresponding augmented Lagrangian function, which is the new objective of the minimization problem, is
$$\mathrm{AL}(\xx)=\sum_{i=1}^m x_i\wt_i+\sum_{i=1}^n\lambda_i\big(1-\sum_{j\in \sigma_i}x_j\big)+
\frac{\mu}{2}\sum_{i=1}^n\big(1-\sum_{j\in \sigma_i}x_j\big)^2$$
$$=\sum_{i=1}^m x_i\wt_i+\sum_{i=1}^n\Big(\lambda_i\big(1-\sum_{j\in \sigma_i}x_j\big) + \frac{\mu}{2}\big(1-\sum_{j\in \sigma_i}x_j\big)^2\Big).$$

Since $$\big(1-\sum_{j\in \sigma_i}x_j\big)^2=1+\sum_{j\in \sigma_i}x_j^2+2\!\!\!\sum_{j<k\in \sigma_{i}}\!\!\!x_jx_k-2\sum_{j\in \sigma_i}x_j,$$
and using that $x_j^2=x_j$ for $x_j\in\{0,1\}$, we get
$$\mathrm{AL}(\xx)= \sum_{i=1}^m x_i\wt_i+\sum_{i=1}^n\Big(-\lambda_i\sum_{j\in \sigma_i}x_j+\mu\!\!\!\sum_{j<k\in \sigma_{i}}\!\!\!x_jx_k-\frac{\mu}{2}\sum_{j\in \sigma_i}x_j\Big)+C,$$
\begin{equation}
	= \sum_{i=1}^m x_i\wt_i+\sum_{i=1}^n\Big((-\lambda_i-\frac{\mu}{2})
	\sum_{j\in \sigma_i}x_j+\mu\!\!\!\sum_{j<k\in \sigma_{i}}\!\!\!x_jx_k\Big)+C,\label{eq:QUBO_AL}
\end{equation}
where $C$ is a constant, which can be ignored when solving the optimization problem. From \cref{eq:QUBO_AL}, we can get the coefficients $J_{ij}$ and $h_i$ of the QUBO representation \cref{eq:Ising1} and solve that QUBO on a quantum annealer, updating parameters $\ll$ and $\mu$ at each iteration as specified in \cref{alg:AL}.

\vspace{1cm}
\subsection{HUBO version}\label{sec:hubo}
Higher-order binary optimization (HUBO) is a generalization of quadratic unconstrained binary optimization (QUBO) to higher-order polynomials. Each HUBO problem can be converted into a QUBO problem by defining auxiliary variables that encode products of other variables in a way that leads to decreasing the polynomial degree. For instance, if monomial $x_1x_2x_3$ is part of a HUBO, one can define a new variable $u=x_1x_2$ and replace $x_1x_2x_3$ by $uv_3$. The constraint $u=x_1x_2$ can be enforced using a well-known penalty quadratic function $x_1x_2-2(x_1+x_2)u+3u$ \cite{DW_Manual} resulting into the QUBO
$$ux_3+\mu(x_1x_2-2(x_1+x_2)u+3u),$$
which can replace $x_1x_2x_3$ for sufficiently large penalty $\mu$. Applying this repeatedly can convert HUBO of any degree into a QUBO.

To formulate a HUBO version of the SCP, we start with the 0-1 linear program \cref{eq:SCP1}--\cref{eq:SCP2}. 
The constraint $\sum_{j\in\sigma_i}x_j\geq 1$ means that, for  at least one $j\in \sigma_i$, $x_j=1$, which means that, for at least for one $j\in \sigma_i$, $1-x_j=0$.
Hence, the constraints \cref{eq:SCP2} are equivalent to
$$\prod_{j\in \sigma_i}(1-x_j)=0, \;  i=1,...,n.$$
and
\begin{equation}
	\sum_{i=1}^n \prod_{j\in \sigma_i}(1-x_j)=0. \label{eq:product}
\end{equation}
Define new binary variables $y_i=1-x_i$, $i=1,\dots,m$. Replacing $x_j$ in \cref{eq:product} and \cref{eq:SCP1} with $y_j$ and combining them into a single function, we get the HUBO formulation of the SCP
\begin{equation} 
	\mbox{minimize}\bigg(m-\sum_{i=1}^my_i\wt_i+\mu\sum_{i=1}^n \prod_{j\in \sigma_i}y_j\bigg), \label{eq_final_hubo}
\end{equation} 
where $\mu$ is a penalty coefficient. Clearly, the constant $m$ can be ignored for the optimization. 

It is easy to see that it is enough to choose $\mu>\max\{\wt_i\}$. Assume that $\mu>\max\{\wt_i\}$ and, in a solution $\{y_i\}$ of \cref{eq_final_hubo}, there is an element $i$ that is not covered, i.e., $\prod_{j\in \sigma_i}y_j=1$.  Then there will exist a set $S_i$ that contains $j$ and is not in the cover, i.e., $y_i=1$ (since by the SCP definition $\bigcup_i S_i=U$). Adding $S_i$ to the cover, i.e., changing $y_i$ to 0, will change the value of the objective function by $\wt_i-\mu<0$, which contradicts the assumption that $\{y_i\}$ is an optimal solution of \cref{eq_final_hubo}.

\section{Results} \label{sec:results}
\subsection{Implementation of the algorithms}
For our experiments, we use the D-Wave \texttt{Advantage\_system4.1} quantum annealer, which we call \textit{DWA} hereafter, available through the Leap quantum cloud service. 
The annealing parameters we use to control the annealing process are $\texttt{num\_reads}=1000$ 
for the number of samples returned per call to the annealer, $\texttt{annealing\_time}=100$, 
which sets the number of microseconds for the annealing time, and $\texttt{chain\_break\_method}=\texttt{MinimizeEnergy}$. 
We also use the \texttt{flux\_biases} parameter, which is used to help control some hardware biases.
Unless explicitly stated otherwise, all other parameters are set to their default values. 

We test the proposed new algorithms and compare them against the standard approach on random instances of the SCP. To generate the test problems, our generator for SCP instances takes as an input the number of sets $m$, the number of elements $n$, and the \textit{coverage} $c$, defined as the average number of sets covering an element of $U=\{1,\dots,n\}$. The generator initially creates $m$ empty sets and then randomly places elements in sets $S_i$ until the following conditions are satisfied at completion: (i) each element of $U$ is contained in at least two sets, (ii) each set contains at least one element, and (iii)  $\sum_i|S_i|\geq mc$. Finally, a random weight $\wt_i\in[0,1]$ for $i=1,\dots,m$ is assigned to each set $S_i$. We use different size parameters for $m$ in $\{50,75,100,\dots,400\}$ and, for each $m$, we compute a set ${N}_m$ of three values for $n$ defined as ${N}_m=\{\lceil{0.5m}\rceil,\,\lceil{0.75m}\rceil,\, m\}$. For the value of coverage we use $c=3$. For each type of experiment and combination of $m$ and $n$, we generate three random instances and average the values of the measured characteristic over the three instances. If the annealer returns an infeasible solution, i.e., one that doesn't correspond to a valid cover, we define as cost of that solution the sum of all weights, which corresponds to the trivial cover consisting of all sets.

\begin{table*}[]
	\begin{tabular}{|c|c|c|c|}
		\hline
		Name     & Inequalities method  & Optimization method & Classical/quantum \\ \hline\hline
		\texttt{SV\_QA}   & slack variable       & quantum annealing   & quantum           \\ \hline
		\texttt{AL\_SA}   & augmented Lagrangian & simulated annealing & classical         \\ \hline
		\texttt{AL\_QA}   & augmented Lagrangian & quantum annealing   & quantum           \\ \hline
		\texttt{HUBO\_SA} & HUBO                 & simulated annealing & classical         \\ \hline
		\texttt{HUBO\_QA} & HUBO                 & quantum annealing   & quantum           \\ \hline
	\end{tabular}
	\caption{Implemented algorithms used for the experimental analysis.}\label{tab:algs}
\end{table*}

The algorithms used in the experiments are the following (see \cref{tab:algs}). 

\texttt{SV\_QA}: based on the standard slack-variable formulation from \cref{sec:slack} and implementation on a quantum annealer. 

\texttt{AL\_SA}: uses the augmented Lagrangian representation as a model and the simulated annealing method to do the optimization. \textit{Simulated annealing} \cite{van1987simulated} is a classical optimization method that uses a probabilistic approach to explore the solution space in search for a global minimum. To escape local minima, it uses gradual cooling from a high temperature to a low one,  which changes the probability of accepting moves increasing the objective function from higher in the beginning to low at the end. 
We have opted for simulated annealing in our experiments because it serves as a classical analogue of quantum annealing being a general-purpose heuristic for global optimization. To carry out our experiments, we utilized the implementation of simulated annealing that is included in the Ocean D-Wave software. The ALM parameters from \cref{alg:AL} are $0.5$ for the initial $\mu$, $0$ for the initial $\lambda$, and $1.1$ for the increase factor $\rho$.

\texttt{AL\_QA}: the quantum version of \texttt{AL\_SA},combining ALM with quantum annealing. We use the same ALM parameters as \texttt{AL\_SA}.

\texttt{HUBO\_SA}: uses the HUBO formulation from \cref{sec:hubo}. For solving the HUBO we use the simulated annealing solver of Ocean. Note that, since simulated annealing is not restricted to quadratic models, we don't need a conversion to a QUBO in this algorithm.

\texttt{HUBO\_QA}: the quantum version of \texttt{HUBO\_SA} as it uses the HUBO formulation of the SCP. But in order to solve on DWA, we also need a HUBO-to-QUBO converter, for which we use the one supplied by the Ocean software.

In the next three subsections, we will analyze the implementations of the augmented Lagrangian and HUBO algorithms and compare all five algorithms with respect to the qualities of the proposed solutions to SCP.

\subsection{Augmented Lagrangian method iterations}
\begin{figure*}
	\centering
	\includegraphics[width=0.99\textwidth]{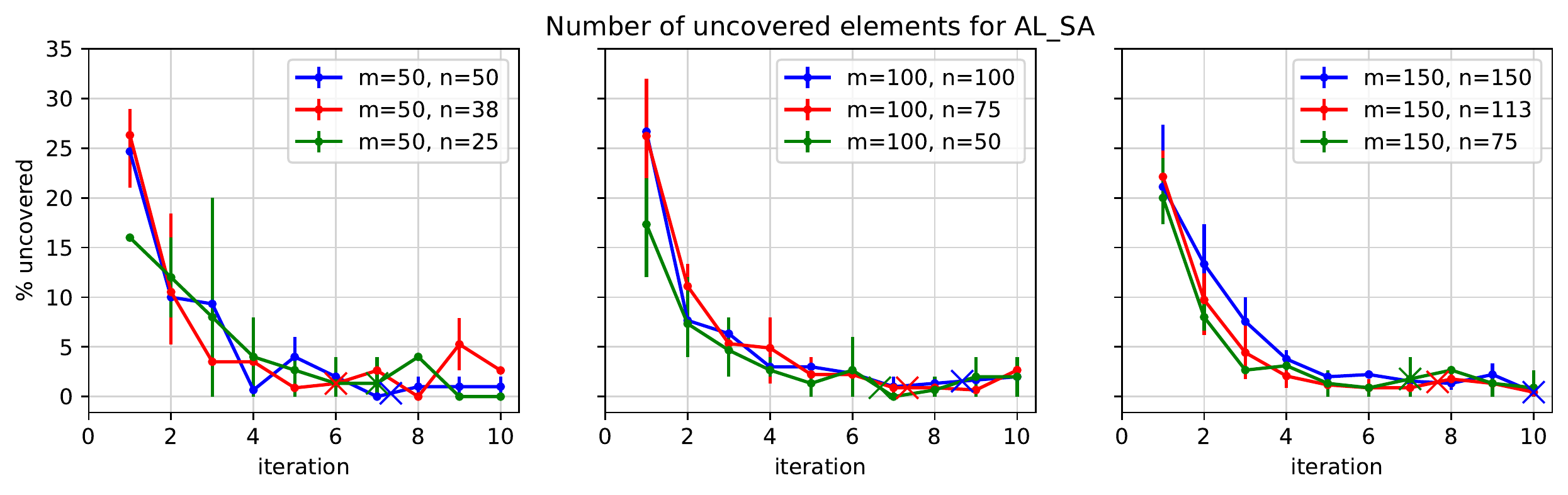}\\ [1.5ex]
	\includegraphics[width=0.99\textwidth]{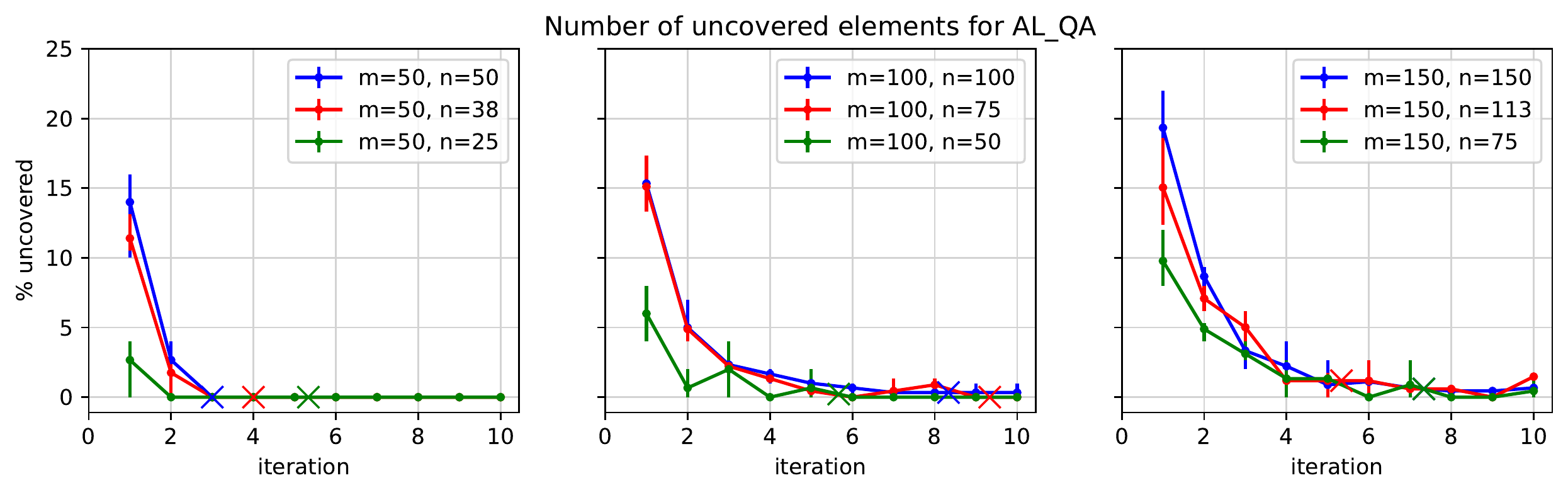}
	\caption{Number of elements not covered, per iteration step, for $m\in\{50,100,150\}$. Shown at the top is \texttt{AL\_SA} and at the bottom it is \texttt{AL\_QA}. A ``{$\times$}'' symbol indicates the average iteration number at which the best solution for the corresponding problem size combination has been obtained.}\label{fig:AL_iters}
\end{figure*}

Our focus here is on analyzing the iterations of ALM, specifically, examining the decrease rate in the number of non-satisfied constraints as the iteration number increases. \cref{fig:AL_iters} shows the percentage of uncovered elements (non-satisfied constraints), for \texttt{AL\_SA} and \texttt{AL\_QA} and for number of variables $m\in\{50,100,150\}$. The number of ALM iterations is set to 10. We observe that, in all cases, for smaller values for $n$ we see a better performance. Although this may seem self-evident, we will see in the next subsection that decreasing the value of $n$ makes the problem harder in other aspects. Regarding the dependence on the value of $m$, we see for \texttt{AL\_QA} that with increasing the value of $m$ the percentage of uncovered elements also goes up. We cannot see such a clear trend for \texttt{AL\_SA}. Comparing \texttt{AL\_SA} and \texttt{AL\_QA} implementations, we see that, with respect to this criterion, \texttt{AL\_QA} has a better performance.

Finally, we look at the value of the penalty factor $\mu$ from \cref{alg:AL}. The magnitude of $\mu$ is important, especially in the case of quantum annealing, since it is used to scale up some coefficients, and large values of $\mu$ can negatively affect the accuracy of the annealing, as discussed earlier. While \cref{fig:AL_iters} doesn't directly show these values, they can be easily calculated given the iteration number $i$, e.g., in our implementation, $\mu(i)=0.5(1.1)^i$. \cref{fig:AL_iters} shows the average iteration number where the best solution was found. 
We observe that, in the case $m=50$, the best-iteration numbers and the corresponding values of $\mu$ are lower compared to $m=100$ and $m=150$, especially for algorithm  \texttt{AL\_QA}. However, we don't see significant difference when we compare $m=100$ and $m=150$. One possible explanation is that the number of iterations, ten, may be not enough for some instances with large values of $m$ and the best iteration number for $m=150$ may be greater than 10. But increasing the number of iterations also increases the cost of the algorithm.

\subsection{Number of variables and embeddability}
The computational complexity of a classical optimization algorithm goes up with the number of variables of the instance. For quantum algorithms, larger number of variables usually means lower quality of the solution. But a large number of variables may also means that the QUBO does not fit on the quantum device, so the problem may not be solvable at all. Two key factors influence the sizes of problems that can be solved on a quantum computing device: the number of qubits and the connections between them. The DWA device has more than 5000 qubits, with each qubit connected to no more than 15 other qubits. The sparsity of connections means that, for most problems, several connected qubits have to be combined into a single logical qubit to represents one binary variable. Hence, the sizes of the problems that can be solved on DWA, in terms of the number of binary variables, may be much smaller than the number of available qubits. In this section, we analyze what sizes of problems are solvable for each of the three quantum algorithms: \texttt{SV\_QA}, \texttt{AL\_QA}, and \texttt{HUBO\_QA}.

First, let us analyze the number of the variables. \texttt{SV\_QA} uses a QUBO formulation with $m+n(\log m+1)$ \cite{Lucas2014}. \texttt{AL\_QA} also uses a QUBO formulation and its number of binary variables is $m$. Finally, \texttt{HUBO\_QA} uses a HUBO formulation with $m$ variables. But in order to convert it to a quadratic form, one should define a number of auxiliary variables to reduce the degree of the monomials, so the final number of variables is higher. The number of auxiliary variables depends on the implementation, and we don't have a formula for the one used by the Ocean software. But we can experimentally analyze them by counting the number of final variables for each test instance. 

\begin{figure*}
	\vspace*{-0.5cm}
	\centering
	\includegraphics[width=0.78\textwidth]{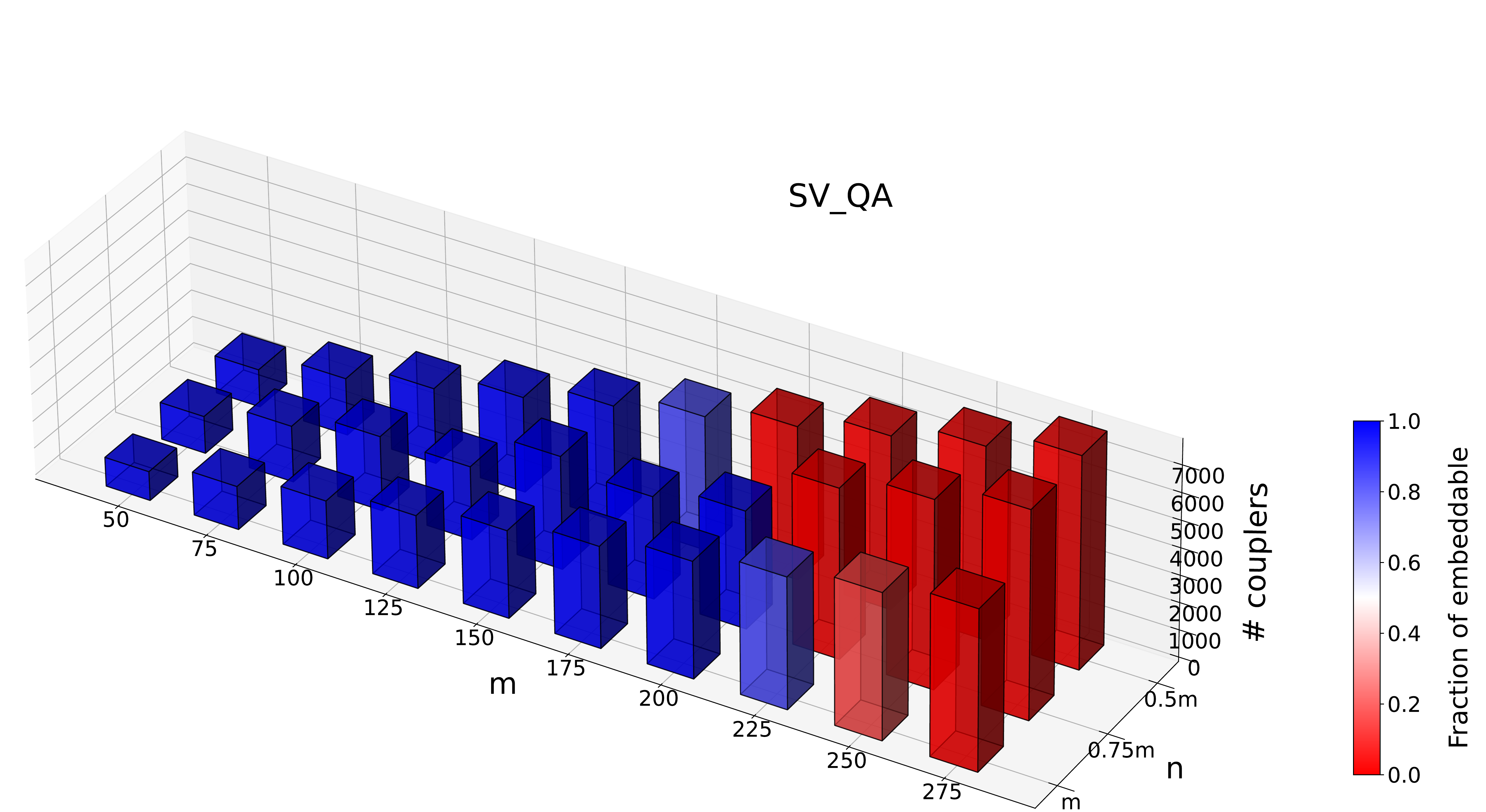}\\ 
	\includegraphics[width=0.78\textwidth]{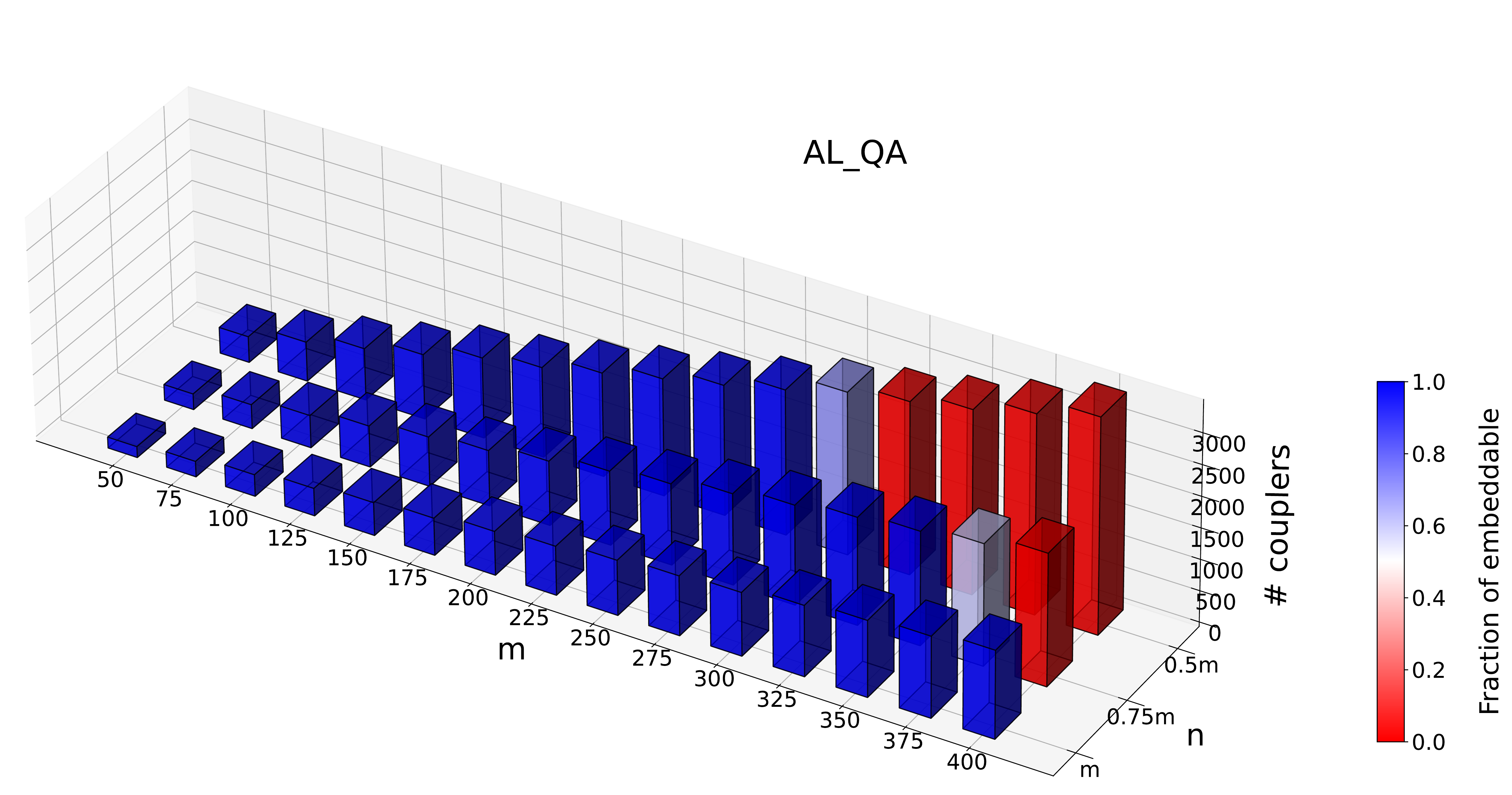}\\ 
	\includegraphics[width=0.78\textwidth]{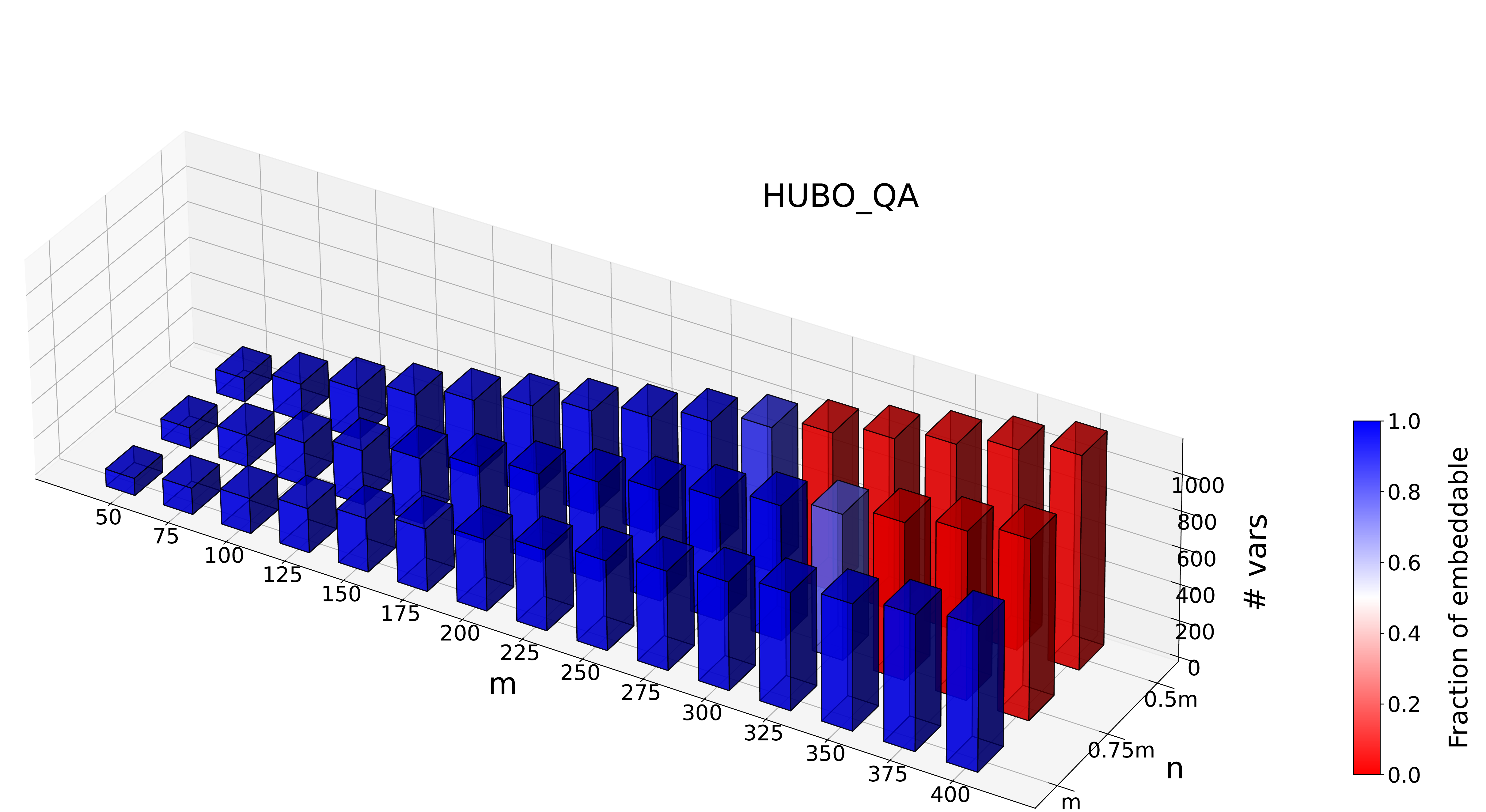}
	\caption{Comparison of the three quantum algorithms with respect to the number of variables or couplers. On the top is \texttt{SV\_QA}, in the middle it is  \texttt{AL\_QA}, and at the bottom it is \texttt{HUBO\_QA}. The colors show what fraction the problem instances are embeddable into the DWA chip.}\label{fig:num_bin_vars}
\end{figure*}

\cref{fig:num_bin_vars}, bottom, shows the number of binary variables for \texttt{HUBO\_QA} for different SCP sizes. For each value of $m$, we observe that the number of binary variables increases with decreasing the value of $n$. While this might look counter-intuitive, it is based on the fact that the number of auxiliary variables depends on the degrees of the monomials in the HUBO, as each monomial of degree larger than two needs additional variables to quadratize it.
Furthermore, the degree of the $i$-th monomial is the size of the set $\sigma_i$ (see \cref{eq_final_hubo}), which is the number of all sets containing element $i$. Smaller values of $n$ means more sets covering each element, on average and, hence, higher degree monomials. We can also see that the number of variables for \texttt{HUBO\_QA} grows when $m$ is varied between 50 an 400 from 95.5 to 794.2, for $n=m$, and from 133.6  to 1170.4, for $n=m/2$.

\cref{fig:num_bin_vars}, top and middle, displays the number of couplers (quadratic coefficients) for  \texttt{SV\_QA} and \texttt{AL\_QA}, respectively. We don't plot the number of QUBO variables for these algorithms as they can be calculated using explicit formulas, which are $m+n(\log m+1)$ and $m$.
The number of couplers is important since, without all-to-all connectivity, larger number of couplers usually means that more qubits are needed to represent a single QUBO variable, which reduces the sizes of the embeddable problems. 
We observe a similar pattern, which is that the number of couplers increases when $m$ goes up or $n$ goes down. Specifically, when $n$ decreases, the average size of $\sigma_i$ increases, while the numbers of couplers for  $i$ goes up as roughly $(|\sigma_i|+\log m)^2$ for \texttt{SV\_QA} and $|\sigma_i|$ for \texttt{AL\_QA}, see \cref{eq:SV} and \cref{eq:QUBO_AL}. Specifically, for $m=275$ and $n=138$, the number of couplers for \texttt{SV\_QA} is 7794 and for \texttt{AL\_QA} it is 2326.

Whether a particular instance of the SCP can be embedded in the DWA chip depends on both the \textit{number} of variables and the number of couplers, but it also depends on the specific \textit{connection patterns}, e.g., the locality of the connections. Since the connection patterns are hard to quantify, the ultimate criterion for evaluating and comparing the algorithms we use is the likelihood for problems with given values of $n$ and $m$ to be embeddable. On \cref{fig:num_bin_vars}, the colors  indicate what portion of the ten random problems generated for each specific combination of $n$ and $m$ could be successfully embedded. We can observe that, for $m$ upto 150, all algorithms produce embeddable QUBOs 100\% of the time. At $m=175$, \texttt{SV\_QA} could still embed all problems, except one instance for $n=138$, and the other two algorithms can embed all variables 100\% of the time. Despite that single infeasible instance, we take $m=175$ to be the largest value for $m$ where all methods are able to produce solutions, and in the next subsection we compare the methods with respect to their accuracy for $m$ upto 175. 

For values of $m$ between $200$ and $275$, all instances of \texttt{AL\_QA} and \texttt{HUBO\_QA} are embeddable, while none of the \texttt{SV\_QA} instances are. Although \texttt{HUBO\_QA} has more embeddable instances for $m\geq 300$, both methods have a 33\% embeddability rate for $m=400$, and for $m=425$, no instances are embeddable. 
It is noteworthy that \texttt{HUBO\_QA} performs just slightly worse than \texttt{AL\_QA} with respect to embeddability in DWA, despite having a substantially higher number of QUBO variables (ranging from two to three times more, depending on $n$).

\subsection{Solution quality comparison}
\begin{figure*}
	\centering
	\includegraphics[width=0.99\textwidth]{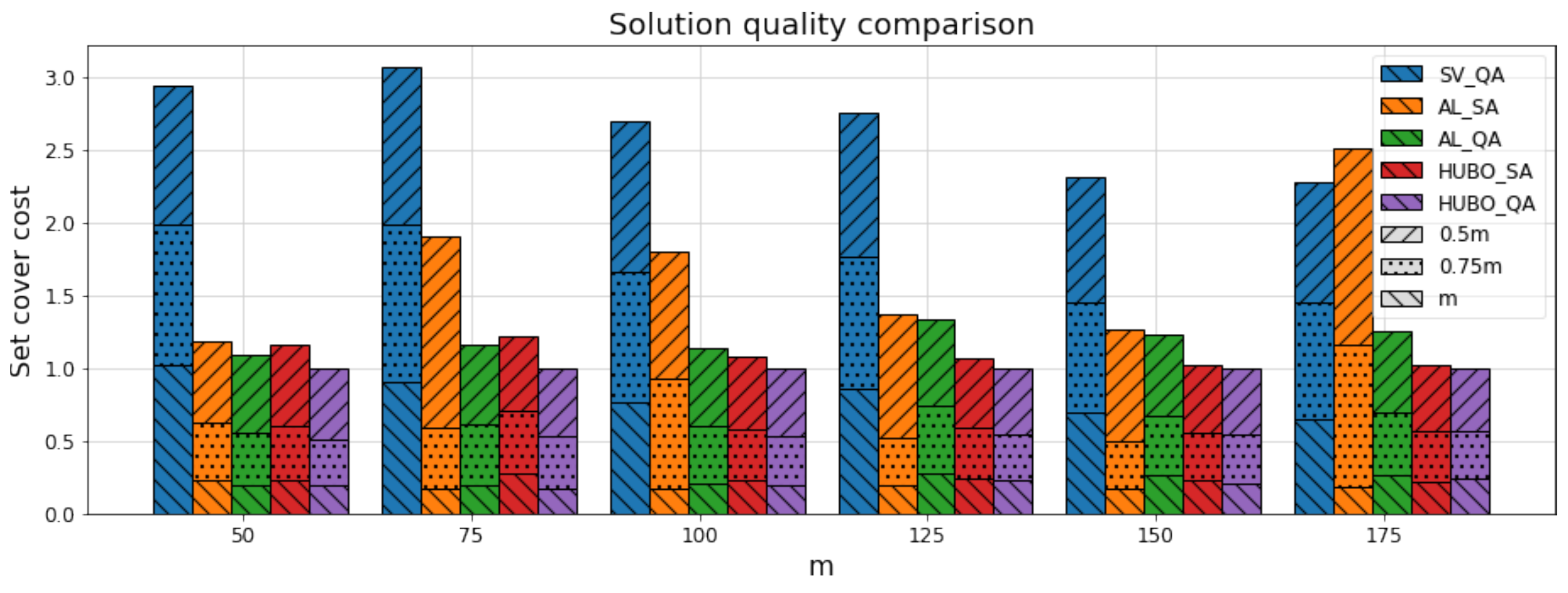}
	\caption{Comparison of the methods based on the set cover costs of computed solutions. Each method and combination of $m\in\{50,75,\dots,175\}$ is represented by a bar, with shading indicating a different value of $n$.}\label{fig:cost_comparison}
\end{figure*}

\cref{fig:cost_comparison} shows the average costs of the covers computed by the five algorithms discussed in this paper for different values of $m$ and $n$. To make the comparisons easier, the weights on the sets for each value of $m$ have been renormalized so that the bar height for \texttt{HUBO\_QA} is one.
Overall, in terms of solution quality, \texttt{HUBO\_QA} performs the best, while \texttt{SV\_QA} performs the worst. To provide a more detailed analysis, we compare the performance of quantum and classical algorithms, followed by a comparison of the algorithms based on the optimization method used..

Our analysis compares the performance of quantum and classical versions of our proposed methods, specifically  \texttt{AL\_QA} vs. \texttt{AL\_SA} and \texttt{HUBO\_QA} vs. \texttt{HUBO\_SA}. We found that quantum annealing (QA) consistently outperforms simulated annealing (SA), particularly in the most challenging case where $n=\lceil m/2 \rceil$ and when comparing \texttt{AL\_QA} with \texttt{AL\_SA}. On the other had, \texttt{HUBO\_SA} is just slightly worse than \texttt{HUBO\_QA}. A possible explanation is that, while the \texttt{HUBO\_QA} performance is degraded with increasing the size of the problem due to the QUBO-to-HUBO conversion, our implementation of \texttt{HUBO\_SA} directly applies simulated annealing to the HUBO representation and does not suffer from an increased number of variables.

Finally, let us compare the algorithms with respect to the methods used to implement the optimization constraints: slack variables (\texttt{SV\_QA}), augmented Lagrangian (\texttt{AL\_SA} and \texttt{AL\_QA}), and HUBO (\texttt{HUBO\_QA} and \texttt{HUBO\_SA}). The slack variables approach finds worse solutions than the others. Except in one case, $m=175$, where \texttt{AL\_SA} is a bit worse than, in the other cases the solutions found by \texttt{SV\_QA} have cost about twice greater on average than the costs of the solutions produced by the other methods. 
In contrast to the other methods, the performance of \texttt{SV\_QA} is relatively consistent across different values of $n$ when $m$ is fixed, resulting in little variation in the quality of the solutions obtained. The HUBO method  performed the best, with both the QA and SA implementations finding better solutions than the other methods. ALM, in particular its QA version, performed slightly worse, but seems to be a viable alternative.

\section{Conclusion}
This paper focuses on the set cover problem and the challenge of implementing multiple inequality constraints on a quantum annealer. We compare the standard approach based on slack variables \cite{Lucas2014} with two new approaches based on the augmented Lagrangian method (ALM) and higher-order binary optimization (HUBO), respectively. Our experimental analysis shows that both new approaches outperform the standard approach. The HUBO approach finds the highest quality solutions and is easy to implement. However, unlike ALM, which is more general and can be applied straightforwardly to other problems with inequality constraints, the HUBO formulation relies on the specific structure of the set cover problem and may not be applicable to other problems with inequality constraints. Also, the HUBO approach is less scalable than ALM in terms of embeddability in the quantum chip of D-Wave.

We also demonstrate that even with a large number of inequality constraints, the augmented Lagrangian method may be a viable approach for solving constrained problems on a quantum annealer. We perform experiments with problems having up to $400$ constraints and find good quality solutions. However, the quality of solutions tends to degrade slowly with increasing $n$ and $m$, but it is much more affected by the number of quadratic couplers of the QUBO, which is determined by the ratio of $n/m$.

Our results could be applicable to solving other optimization problems with constraints on quantum annealers. Future research directions could include improving the implementation of the augmented Lagrangian optimization procedure (Algorithm \ref{alg:AL}), for instance, by updating the stopping criterion or the method for updating the values of $\mu$ and $\ll$. Additionally, the conversion from HUBO to QUBO could be improved to produce fewer auxiliary variables. The Ocean implementation we used is good, but better implementations are possible, especially those that take into account the specific structure of the HUBO.

\section*{Acknowledgments}
\label{sec:acknowledgments}
This work was supported by grant number KP-06-DB-11 of the Bulgarian National Science Fund and by the Laboratory Directed Research and Development program of Los Alamos National Laboratory under project 20210114ER. Los Alamos National Laboratory is operated by Triad National Security, LLC, for the National Nuclear Security Administration of U.S. Department of Energy (contract No.\ 89233218CNA000001).

\bibliographystyle{plain}
\bibliography{quantum}

\end{document}